\documentclass[fleqn,usenatbib]{mnras}
\usepackage{times}
\usepackage{graphicx}
\usepackage{amsfonts,amsmath,amssymb}
\usepackage[usenames,dvipsnames]{color}
\newcommand\herschel{\textit{Herschel}}

\usepackage{hyperref}
\hypersetup{colorlinks=true,pdfborder={0 0 0}}

\newcommand{\intensity}{\mbox{\ensuremath{\mathrm{W}\,\mathrm{m^{-2}}\,\mathrm{Hz^{-1}}\,\mathrm{sr^{-1}}}}}

\newcommand{\etaff}{\mbox{\ensuremath{\eta_\mathrm{ff}}}}
\newcommand{\etadiff}{\mbox{\ensuremath{\eta_\mathrm{diff}}}}

\title[SPIRE-FTS Extended-Source Calibration]{Correcting the extended-source calibration for the \textit{Herschel}-SPIRE Fourier-Transform Spectrometer}

\author[Valtchanov et al.]{I. Valtchanov$^{1}$\thanks{E-mail: ivaltchanov@esa.int},
R. Hopwood$^{1,2}$,
G. Bendo$^{3}$,
C. Benson$^{4}$,
L. Conversi$^{5}$,
T. Fulton$^{4,6}$,\newauthor
M. J. Griffin$^{7}$, 
T. Joubaud$^{2}$,
T. Lim$^{1,8}$,
N. Lu$^{9}$,
N. Marchili$^{10}$,
G. Makiwa$^{4}$,\newauthor
R. A. Meyer$^{2,11}$,
D. A. Naylor$^{4}$,
C. North$^{7}$, 
A. Papageorgiou$^{7}$,
C. Pearson$^{8,12,13}$,\newauthor
E. T. Polehampton$^{8}$, 
J. Scott$^{4}$,
B. Schulz$^{14}$, 
L. D. Spencer$^{4}$,
M. H. D. van der Wiel$^{4,15}$, \newauthor
R. Wu$^{16}$,
\\
$^{1}$Telespazio Vega UK for ESA, European Space Astronomy Centre, Operations Department, 28691 Villanueva de la Ca\~nada, Spain\\
$^{2}$Department of Physics, Imperial College London, Prince Consort Road, London SW7 2AZ, UK \\
$^{3}$UK ALMA Regional Centre Node, Jodrell Bank Centre for Astrophysics, School of Physics and Astronomy, University of Manchester, \\
Manchester M13 9PL, UK\\
$^{4}$Institute for Space Imaging Science, Department~of Physics \& Astronomy, University of Lethbridge, 4401 University Drive, \\
Lethbridge, Alberta, T1K 3M4, Canada \\
$^{5}$European Space Astronomy Centre (ESA/ESAC), Operations Department, 28691 Villanueva de la Ca\~nada, Spain\\
$^{6}$Blue Sky Spectroscopy, Lethbridge, AB, T1J 0N9, Canada\\
$^{7}$School of Physics and Astronomy, Cardiff University, The Parade, Cardiff, CF24  3AA, UK\\
$^{8}$RAL Space, Rutherford Appleton Laboratory, Chilton, Didcot, Oxfordshire, OX11 0QX, UK \\
$^{9}$South American Center for Astronomy, CAS, Universidad de Chile, Camino El Observatorio 1515, Las Condes, Santiago, Chile\\
$^{10}$IAPS-INAF, Via Fosso del Cavaliere 100, I-00133 Roma, Italy\\
$^{11}${Institute of Physics, Laboratory of Astrophysics, Ecole Polytechnique F\'ed\'erale de Lausanne, CH-1015 Lausanne, Switzerland}\\
$^{12}$School of Physical Sciences, The Open University, Milton Keynes, MK7 6AA, UK\\
$^{13}$Oxford Astrophysics, Denys Wilkinson Building, University of Oxford, Keble Rd, Oxford OX1 3RH, UK\\
$^{14}$Infrared Processing and Analysis Center, California Institute of Technology, MS 100-22, Pasadena, CA 91125, USA \\
$^{15}$ASTRON, the Netherlands Institute for Radio Astronomy, Postbus 2, 7990 AA Dwingeloo, The Netherlands\\
$^{16}$LERMA, Observatoire de Paris, PSL Research University, CNRS, Sorbonne Universit\'{e}s, UPMC Univ. Paris 06, 92190 Meudon, France \\
}

\date{\today}

\begin{document}

\date{Accepted.. Received..; in original form \date{\today}}

\pagerange{\pageref{firstpage}--\pageref{lastpage}} \pubyear{2017}

\maketitle 

\label{firstpage}

\begin{abstract}
We describe an update to the \textit{Herschel}-SPIRE Fourier-Transform Spectrometer (FTS) calibration for extended sources, which incorporates a correction for the frequency-dependent far-field feedhorn efficiency, \etaff. This significant correction affects all FTS extended-source calibrated spectra in sparse or mapping mode, regardless of the spectral resolution.  Line fluxes and continuum levels are underestimated by factors of 1.3--2 in the Spectrometer Long-Wavelength band (SLW, 447--1018 GHz; 671--294 \micron) and 1.4--1.5 in the Spectrometer Short-Wavelength band (SSW, 944--1568 GHz; 318--191 \micron). The correction was implemented in the FTS pipeline version 14.1 and has also been described in the SPIRE Handbook since Feb 2017. Studies based on extended-source calibrated spectra produced prior to this pipeline version should be critically reconsidered using the current products available in the Herschel Science Archive. Once the extended-source calibrated spectra are corrected for \etaff, the synthetic photometry and the broadband intensities from SPIRE photometer maps agree within 2-4\% -- similar levels to the comparison of point-source calibrated spectra and photometry from point-source calibrated maps. The two calibration schemes for the FTS are now self-consistent: the conversion between the corrected extended-source and point-source calibrated spectra can be achieved with the beam solid angle and a gain correction that accounts for the diffraction loss.
\end{abstract}


\section{Introduction}

The calibration of an instrument consists of two tasks: (i) removing all instrument signatures from the data and (ii) converting the products to physical units using a suitable calibration schema. For the first task, a good knowledge of the instrument and its response to different conditions (e.g. observing mode, internal and external thermal and radiation environments, the solar aspect angle, etc.) is required. For the second task, a calibration source of assumed flux or temperature is used to covert the measured signal to physically meaningful units. The atmosphere blocks most far-infrared radiation from reaching the ground, therefore the calibration of far-infrared space borne instrumentation requires a bootstrapping approach based on previous observations and theoretical models of candidate sources, typically planets or asteroids. 

An imaging Fourier-Transform Spectrometer (FTS) is part of the Spectral and Photometric Imaging Receiver (SPIRE, \citealt{Griffin_2010}) on board the \herschel\ Space Observatory \citep{Pilbratt_2010}. SPIRE is one of the most rigorously calibrated far-infrared space instruments to date. It underwent five ground-based test campaigns and regular calibration observations during the nearly four years of in-flight operations of \herschel. The stable space environment at the second Lagrange point and the flawless operation of the instrument resulted in unprecedented accuracy both in terms of the telescope and instrument response. A detailed description of the FTS instrument and its calibration scheme is provided in \citet{Swinyard_2010}, with an update in \citet{Swinyard_2014}.

There are no prior systematic studies of the extended-source calibration for the FTS. Extended-source calibrated maps from the SPIRE Photometer, corrected to the absolute zero level derived via cross-calibration with \textit{Planck}-HFI \citep{Bertincourt_2016}, became available during the post-operations phase of \herschel. These maps allowed for a detailed comparison between photometry and spectroscopy of extended sources. Initial checks showed significant and systematic differences at levels of 40-60\% across the three photometer bands. Some authors also reported discrepancies (\citealt{Kohler_2014, Kamenetzky_2014}) and implemented corrections in order to match the spectra with the photometry. Others proceeded by starting from the point-source calibration and correcting for the source size (e.g. \citealt{Wu_2014,Makiwa_2016,Schrim_2017,Kamenetzky_2015, Morris_2017}). 

The reported differences with the photometer did not initially draw our attention, because the comparison is intricate and depends on the assumptions made. As shown in \citet{Wu_2013}, the coupling of sources that are neither point-like nor fully extended (i.e. semi-extended) require good knowledge of the FTS beam and its side-lobes, as well as good knowledge of the source brightness distribution. Even extended sources with significant sub-structure couple in a complicated way with the multi-moded and non-Gaussian beam \citep{Makiwa_2013}. Moreover, the source size would imply colour-correcting the photometry (see The SPIRE Handbook, \citealt{handbook}, section 5.8; H17 from now on). Hence both sides of the comparison need their proper corrections.

In this study, we have tried to alleviate some of the uncertainties by carefully selecting truly extended sources for cross-comparison with broad-band intensities from the SPIRE Photometer extended-source calibrated maps. The results of this analysis show a significant correction is needed in order to match the extended-source calibrated spectra with the photometry. This paper introduces the methods used to derive the necessary corrections, demonstrates the self-consistency between FTS point and extended-source calibrated spectra, and demonstrates a good agreement with broadband photometry from the SPIRE Photometer.

\textit{Herschel}'s two other instruments, the Heterodyne Instrument for the Far Infrared (HIFI, \citealt{hifi}) and the Photodetector Array Camera and Spectrometer (PACS, \citealt{pacs}), share some spectral overlap with the SPIRE FTS. Analysis of a sample of calibration targets has shown an overall agreement of $\pm20\%$ between the SPIRE FTS and HIFI, and discrepancies up to a factor of 1.5--2 for comparisons with PACS (Puga et al, in preparation). Noting that the instantaneous bandwidth of HIFI (2.4 or 4 GHz depending on observing mode and band) is only marginally wider than the instrumental line shape of the SPIRE FTS (1.2 GHz), the overall agreement between HIFI and the SPIRE FTS is acceptable. The spectral overlap between the SPIRE FTS and the PACS spectrometer falls in 194--210\,$\mu$m, which is an area affected by a PACS spectral leak (see \citealt{pacs_cal}). Although we have performed a comparison between instruments for a sample of extended sources, some results were inconclusive and we have not included this work in this paper. 

The structure of the paper is as follows. In \autoref{sec:extcal} we briefly outline the extended-source calibration scheme. In \autoref{sec:xcal} we compare FTS results with photometry from SPIRE maps using a selection of spatially extended sources and derive a correction that matches the known far-field feedhorn efficiency. In \autoref{sec:etadiff} we link the two FTS calibration schemes (i.e., the point source and the corrected extended source schemes) using the beam solid angle and a correction for diffraction loss. Some guidelines on using the corrected spectra are presented in \autoref{sec:practical}. In \autoref{sec:impact} we outline the significance of the correction and the impact on deriving physical conditions if the uncorrected spectra are used. In \autoref{sec:conclusions} we present the conclusions.

As much as possible we follow the notations used in the SPIRE Handbook (H17). Throughout the paper we interchangeably use \textit{intensity} and \textit{surface brightness} as equivalent terms, in units of either [MJy sr$^{-1}$]  or [\intensity]\footnote{1 MJy sr$^{-1} = 10^{-20}\, \intensity$.}. 

\section{Telescope model based extended-source calibration}
\label{sec:extcal}

In the following, we briefly outline the main points in the FTS calibration scheme,  which is presented in greater detail in \citet{Swinyard_2014}. 

As there is no established absolute calibration source for extended emission in the far infrared and sub-mm bands, the \herschel\ telescope itself is used as a primary calibrator for the FTS. The usual sources used from ground, such as the Moon and the big planets (e.g. \citealt{Wilson_2013}), are either too close to the Sun/Earth or too bright for the instrument.  

The SPIRE FTS simultaneously observes two very broad overlapping spectral bands. The signals are recorded with two arrays of hexagonally close packed, feedhorn-coupled, bolometer detectors: the Spectrometer Short Wavelength (SSW) array with 37 bolometers, covering 191--318 \micron\ (1568--944\,GHz) and the Spectrometer Long Wavelength (SLW) array with 19 bolometers, covering 294--671 \micron\ (1018--447\,GHz). The bolometers operate at a  temperature of $\sim300$\,mK, which is achieved with a special $^3$He sorption cooler (see H17 for more details).

Within the FTS, the radiation from the combination of the astronomical source, the telescope, and the instrument\footnote{The instrument contribution enters in the total radiation because of the Mach-Zehnder configuration of the FTS, where a second input port views a internal blackbody source (see H17 for more details).} is split into two beams. A moving mirror introduces an optical path difference between the two beams. The recombination of the beams produces an interferogram on each of the individual feedhorn-coupled bolometers. Hence the recorded signal $V_\mathrm{obs}$ after Fourier transforming the interferograms, can be expressed as
\begin{equation}
V_\mathrm{obs}\left[\mathrm{V\, Hz}^{-1}\right] = R_S I_S + R_\mathrm{tel} M_\mathrm{tel} + R_\mathrm{inst} M_\mathrm{inst}, 
\end{equation}
where $I_S$ is the source intensity, $M_\mathrm{tel}$ and $M_\mathrm{inst}$ are the intensities corresponding to the telescope and the instrument emission models. $R_S$, $R_\mathrm{tel}$ and $R_\mathrm{inst}$ are the relative spectral response functions (RSRF) of the system for the source, the telescope and the instrument, respectively. We assume the instrument and telescope emissions to be fully extended in the beam, and well represented by blackbody functions and $R_S = R_\mathrm{tel}$. The units of $I_S$, $M_\mathrm{tel}$ and $M_\mathrm{inst}$ are [\intensity], therefore the RSRF are in units of [V Hz$^{-1}$/(\intensity)].

The instrument is modelled as a single temperature blackbody, $M_\mathrm{inst} = B(\nu,T_\mathrm{inst})$, where $B(\nu,T)$ is the blackbody Planck function and $T_\mathrm{inst}$ is the temperature of the instrument enclosure in Kelvin (available from housekeeping telemetry). The instrument is usually at $\sim5$\,K and following Wien's displacement law, the peak of the instrument emission is at $\sim600$\,\micron, thus $M_\mathrm{inst}$ is much more significant for the longer-wavelength SLW band than for the SSW band.

The telescope model used in the pipeline is a sum of two blackbody models, one for the primary and one for the secondary mirrors:
\begin{equation}
M_\mathrm{tel} = E_{corr}(t)\,\varepsilon_1\,(1 - \varepsilon_2)\,B(\nu,T_{M1}) + \varepsilon_2\,B(\nu,T_{M2}),
\label{eq:mtelcorr}
\end{equation}
where $\varepsilon_1 = \varepsilon_2 \equiv \varepsilon(\nu)$ is the frequency dependent telescope mirror emissivity, and $T_{M1}$ and $T_{M2}$ are the average temperatures of the primary and secondary mirrors, obtained via telemetry from several thermometers placed at various locations on the mirrors. The emissivity in Eq.~\ref{eq:mtelcorr} was measured for representative mirror samples pre-launch by \citet{Fischer_2004}. For a dusty mirror $\varepsilon$ is of the order of 0.2-0.3 \% in the 200-600 \micron\ band, with large systematic uncertainties. The only measured point in the SPIRE band, at 496 \micron, has $\varepsilon = 0.23^{+0.06}_{-0.12}\ \%$. Based on repeatability analysis of a number of ``dark sky'' observations in \citet{Hopwood_2014}, the model was corrected by a small (sub 1\%) and mission-date dependent adjustment to the emissivity, $E_{corr}(t)$.

During the \herschel\ mission around the second Lagrange point of the Earth-Sun system, the primary mirror temperature $T_{M1}$ was of the order of 88\,K and the secondary mirror $T_{M2}$ was colder by 4--5\,K, i.e. at around 84\,K. Even with the low emissivity the telescope thermal emission is the dominant source of radiation recorded by the detectors. Only a few of the sky sources observed with the SPIRE spectrometer are brighter than $M_\mathrm{tel}$: nearby large planets (Mars, Saturn) and the Galactic centre.

The calibration of the FTS requires the derivation of $R_\mathrm{tel}$, $R_\mathrm{inst}$, $M_\mathrm{tel}$ and $M_\mathrm{inst}$, as we can then recover the source intensity using

\begin{equation}
I_S \left[\intensity\right] = \frac{\left(V_\mathrm{obs} - R_\mathrm{inst} M_\mathrm{inst}\right)}{R_\mathrm{tel}} - M_\mathrm{tel}.
\label{eq:iext}
\end{equation}
Note that all quantities in \autoref{eq:iext} are frequency dependent and derived independently for each FTS band (see \citealt{Fulton_2014}). As the two bands SSW and SLW overlap in 944--1018 GHz, the intensities in this region should match within the uncertainties.

\begin{figure*}
\includegraphics[width=8cm]{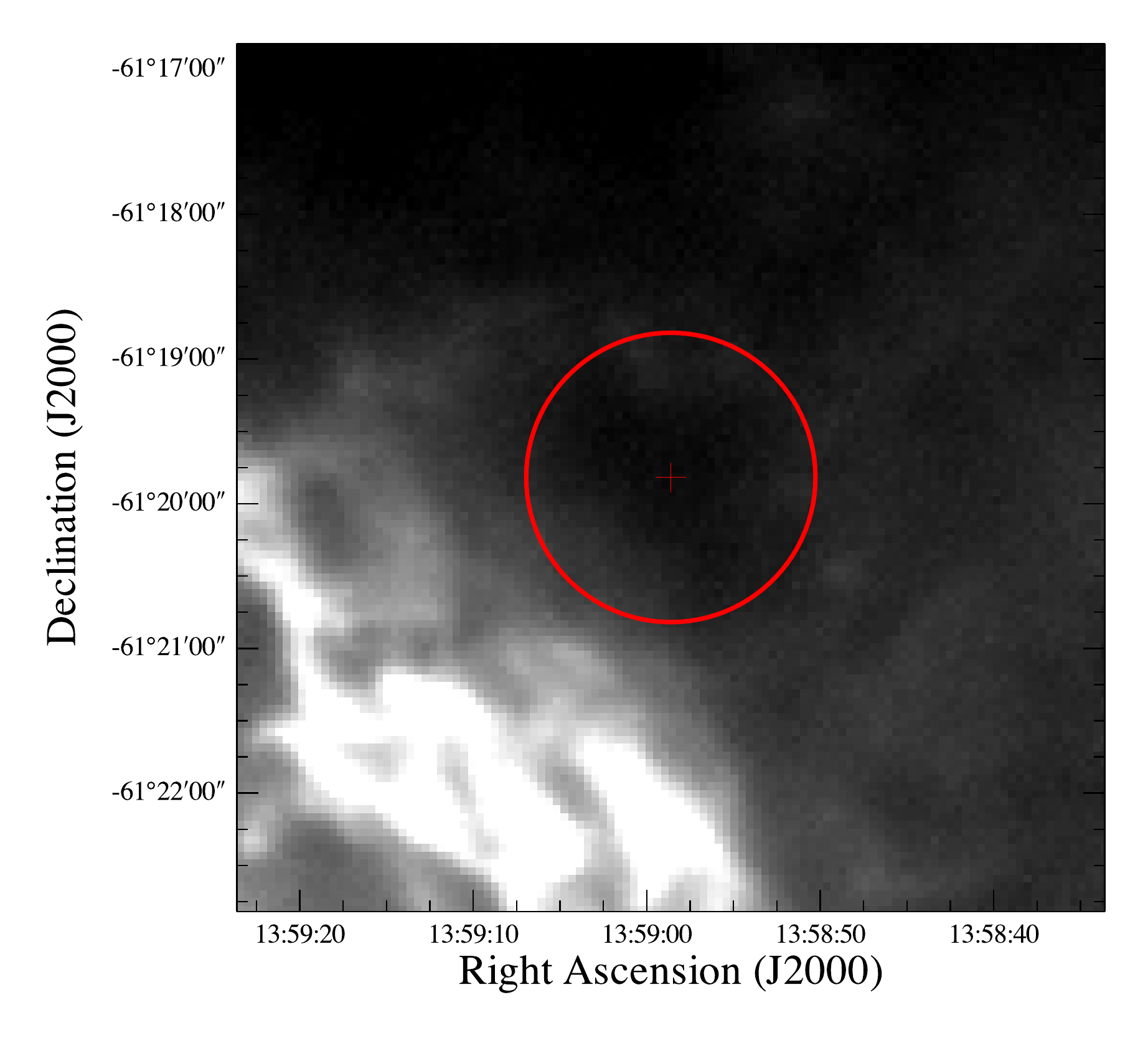}
\includegraphics[width=8cm]{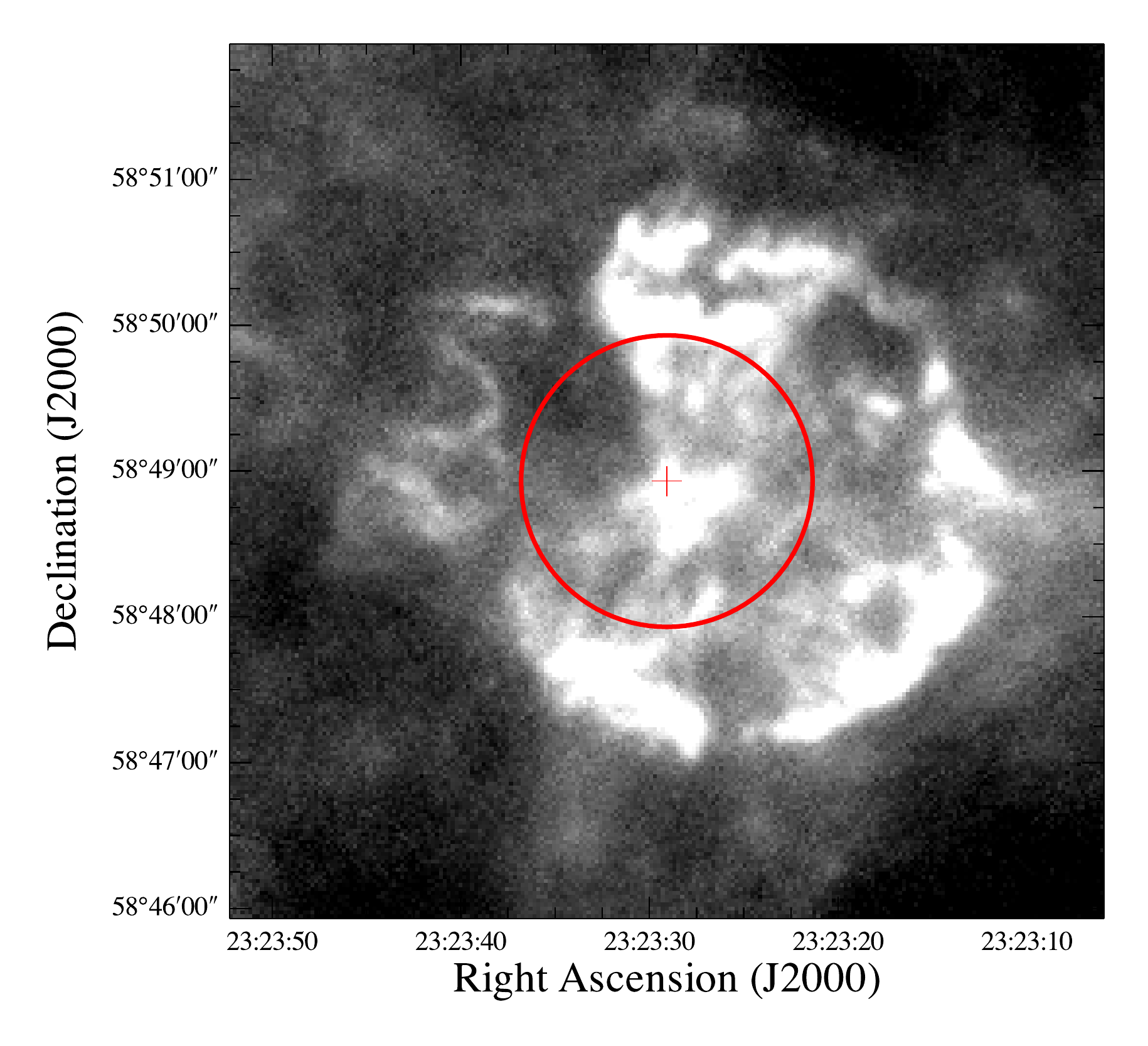}
\caption{(Left) Spatially flat extended source, RCW82off2 (see Table~\ref{tab_obsids}), with greyscale image corresponding to the PACS 70\,\micron\ map and the 1\arcmin\ radius unvignetted FTS field of view shown as a red circle. The centre of the FTS field is marked with a '+' sign. Note that the region appears as dark due to the very bright nearby RCW82; the peak surface brightness within the FTS footprint at 250\,\micron\ is more than 400\,MJy/sr. (Right) Cas A -- a supernova remnant shown with 70\,\micron\ PACS data that was rejected because of its complex morphology although  $\Delta g_\mathrm{max} = 0.07$ and $\sigma_I = 0.07$.}
\label{fig:flat}
\end{figure*}

The point-source calibration is built upon the extended source calibration, using a suitable model of the emission of a point-like source. In the case of the SPIRE FTS, the primary calibrator is Uranus, which has an almost featureless spectrum in the FTS bands and a disk-averaged brightness temperature model known with uncertainties within $\pm 3\%$ (ESA-4 model, \citealt{Moreno_2010, Orton_2014}). The point-source conversion factor, $C_\mathrm{point}$, is derived as $C_\mathrm{point} = M_\mathrm{Uranus}/I_\mathrm{Uranus}$,
where $I_\mathrm{Uranus}$ is the observed extended-source calibration intensity from the planet (following Eq.~\ref{eq:iext}) and $M_\mathrm{Uranus}$ is the planet's model. $M_\mathrm{Uranus}$ is converted from the disk-averaged brightness temperature model in units of K to units of Jy, using the planet's solid angle, as seen from the \textit{Herschel} telescope at a particular observing epoch (see H17 for details). Hence, $C_\mathrm{point}$ is in units of $\left[\mathrm{Jy}/\left(\intensity\right)\right]$. It is important to emphasise that as long as the model $M_\mathrm{Uranus}$ is a good representation of the planet's emission in the FTS bands, then the point-source calibration is invariant with respect to the extended-source calibration.

The point-source calibration was validated using Uranus and Neptune models, which showed an agreement within 3-5\% \citep{Swinyard_2014}. Furthermore, the calibration accuracy was confirmed using a number of secondary calibrators (stars, asteroids) with the agreement at a level of 3-5\% between point-source calibrated spectra  and the photometry from SPIRE point-source calibrated maps \citep{Hopwood_2015}. Therefore we consider the point-source calibration as well established and in this paper our focus is on the extended source calibration.

\section{Cross-calibration with SPIRE Photometer}
\label{sec:xcal}

The SPIRE Photometer and the FTS are calibrated independently and it is therefore important to cross-match measurements from observations of the same target. The cross-calibration can be considered as a critical validation of the different calibrations and whether their derived accuracies could be considered realistic. The cross-calibration in the case of point sources was already mentioned in the previous section, while in this section we restrict our discussion to the extended-source case.

The cross-calibration is performed between the extended-source calibrated spectra, obtained as described in \autoref{sec:extcal}, and the extended-source calibrated SPIRE Photometer maps. These maps use detector timelines calibrated to the integrated signal of Neptune \citep{Bendo_2013} instead of the Neptune peak signal used for point-source calibrated maps. The arbitrary zero-level of each map is matched to the absolute zero level derived from \textit{Planck} \citep{Bertincourt_2016}. There is a good overlap of the SPIRE 350\,\micron\ band with the \textit{Planck}-HFI 857\,GHz band, and a relatively good overlap between the SPIRE 500\,\micron\ band and the \textit{Planck}-HFI 545\,GHz band. There is no \textit{Planck} overlap for the SPIRE 250\,\micron\ band, so an extrapolation is used, based on a modified blackbody curve and the observed SPIRE 250\,\micron\ and \textit{Planck}-HFI intensities (see H17 for more details). The overall uncertainty in the \textit{Planck}-derived zero level is estimated at $\sim10\%$, but for maps that are comparable in size to the \textit{Planck}-HFI beam  (FWHM $\approx 5$\arcmin, \citealt{planck}) the uncertainty can be larger.

One of the most critical ingredients for extended-source calibration for any particular instrument is the knowledge of the beam and how the beam couples to a source (e.g. \citealt{Ulich_1976, Wilson_2013}). Uncertainties on the beam solid angle or the beam profile as a function of frequency will lead to uncertainties in the derived quantities. 

The SPIRE Photometer beam maps were obtained using special observations of fine scans over Neptune and the same region of the sky at a different epoch when Neptune was no longer in the field of view (i.e., the ``shadow'' observation). Thanks to these two observations the photometer beams for the three bands have been characterised out to 700\arcsec\ and the beam solid angles are known down to the percentage level. Analysis of the beam maps for the three photometer bands indicates that the broadband beams are unimodal and their cores are well-modelled with 2-D Gaussians (see H17 and Schultz et al. in preparation).

On the other hand, the FTS beam was only measured out to a radial distance of $45\arcsec$. The beam is multi-moded and far from Gaussian, especially in the SLW band, which exhibits appreciable frequency-dependent beam FWHM variations \citep{Makiwa_2013}. Hence, for sources with significant spatial brightness variation, the coupling with the beam is rather uncertain. Consequently, for the cross-calibration analysis, we need to identify spatially flat sources with as little source structure as possible within the FTS beam.

\subsection{Selecting targets for cross-calibration}
\label{sec:flats}

For all 1825 FTS observations performed with nominal bias mode (sparse and mapping modes, see H17), we extract an $11 \times 11$ pixel ($66\arcsec \times 66\arcsec$) sub-image from the SPIRE 250\,\micron\ photometer map\footnote{Very few FTS observations have no associated SPIRE photometer map.}, centred on the SSW central detector coordinates. The SPIRE 250\,\micron\ beam FWHM is 18\arcsec\ and the largest SPIRE FTS beam has a FWHM of 42\arcsec\ \citep{Makiwa_2013}, so the selected sub-image is bigger than the largest FTS beam FWHM for all frequencies.  To characterise the surface brightness distribution in each sub-image we introduce the relative variation $\sigma_I = \sigma (I_{250})/\bar I_{250}$, where $\sigma (I_{250})$ is the standard deviation of the broadband 250 \micron\ brightness distribution in the region of interest and $\bar I_{250}$ is the average level. Because of the \textit{Planck} zero level normalisation $\bar I_{250} \gg 0$, no zero division effects are expected. To estimate the source flatness we extract the central row and column from the sub-image and calculate two arrays of ratios: North-South : East-West and North-South : West-East. While either ratio alone can identify a vertical or horizontal gradient, the two ratios are needed to detect sources with diagonal gradients. The measure of the maximum gradient $g_\mathrm{max}$ is the maximum value within the two ratio arrays, with $\Delta g_\mathrm{max} = |1 - g_\mathrm{max}|$. We empirically classify a source as flat if $\sigma_I \leq 0.10$ and $\Delta g_\mathrm{max} \leq 0.2$. 

Out of the 1825 FTS observations in nominal mode we identified 70 flat sources observed at high spectral resolution (HR)\footnote{We do not include Low Resolution (LR) observations as in some cases the calibration introduces significant artefacts, mostly in the SLW band \citep{Marchili_2017}.}. Some  are faint, which introduces a large scatter, especially at 500 \micron; hence we only consider those 53 flat HR-mode sources with $\bar I_{250} \geq 100$ MJy/sr.

Furthermore, all of these 53 sources have \textit{Herschel} PACS photometer observations at 160 \micron\ and either at 70 or 100\,\micron. We use the higher angular resolution PACS maps at 70 \micron\ (or 100\,\micron), with the FWHM of the point-spread function of the order of 6--8\arcsec, to visually identify sources which are either point-like, semi-extended or have a significant sub-structure within a region of radius 1\arcmin. As a result of this visual check, we retain 24 out of the 53 sources as our final sample of flat sources. These sources are listed in Table~\ref{tab_obsids}, while \autoref{fig:flat} shows examples of 70\,\micron\ maps for two observations, a source from our selection (left) and a source that was rejected as having a complicated morphology (right).

\begin{table*}
\caption{List of the final selection of spatially flat sources. The target name is that provided by the proposer. The equatorial coordinates RA, Dec are for the central detector from the SSW array. For mapping we only used one FTS sparse snapshot out of 4 or 16 that were used to build the spectral cube. Only one SPIRE Photometer and one PACS Photometer OBSID are provided, although there can be multiple overlapping observations. If PACS and SPIRE Photometer OBSIDs are the same then the observation was taken in Parallel Mode (see H17).}
\begin{tabular}{rlrrllll}
\hline\hline
ID & Target & RA J2000 (deg) & Dec J2000 (deg) & FTS ObsID & Obs Mode & SPIRE Phot ObsID & PACS Phot ObsID \\ \hline
1 & s104off & 304.54185 &  36.77219 & 1342188192 & sparse & 1342244191 & 1342244191 \\
2 & rcw120rhII & 258.10234 & -38.45376 & 1342191230 & sparse & 1342204101 & 1342216586 \\
3 & rcw120off & 258.25602 & -38.45335 & 1342191233 & sparse & 1342204101 & 1342216586 \\
4 & Cas A FTS Centre-1 & 350.87116 &  58.81551 & 1342202265 & sparse & 1342188182 & 1342188207 \\
5 & rho\_oph\_fts\_off & 246.45504 & -24.33656 & 1342204893 & mapping & 1342205094 & 1342238817 \\
6 & rho\_oph\_fts\_off\_2 & 246.43947 & -24.35357 & 1342204894 & mapping & 1342205094 & 1342238817 \\
7 & EL29\_int & 246.81833 & -24.58734 & 1342204896 & sparse & 1342205094 & 1342238817 \\
8 & rcw82off2 & 209.74421 & -61.33031 & 1342204901 & sparse & 1342203279 & 1342203279 \\
9 & rcw82pdr & 209.75750 & -61.42321 & 1342204902 & sparse & 1342203279 & 1342203279 \\
10 & rcw82rhII & 209.86946 & -61.38302 & 1342204904 & sparse & 1342203279 & 1342203279 \\
11 & rcw82off & 210.05859 & -61.41489 & 1342204910 & sparse & 1342203279 & 1342203279 \\
12 & rcw79rHII & 205.09185 & -61.74105 & 1342204913 & sparse & 1342203086 & 1342258817 \\
13 & rcw79off & 205.37508 & -61.77444 & 1342204917 & sparse & 1342203086 & 1342258817 \\
14 & n2023\_fts\_2 &  85.40126 &  -2.22890 & 1342204922 & mapping & 1342215985 & 1342228914 \\
15 & 02532+6028 &  44.30356 &  60.67048 & 1342204928 & sparse & 1342226655 & 1342226620 \\
16 & IRAx04191\_int &  65.51420 &  15.48075 & 1342214851 & sparse & 1342190615 & 1342241875 \\
17 & los\_30+3 & 278.85175 &  -1.23758 & 1342216894 & sparse & 1342206696 & 1342228961 \\
18 & los\_28.6+0.83 & 280.13751 &  -3.48752 & 1342216895 & sparse & 1342218695 & 1342218695 \\
19 & los\_26.46+0.09 & 279.81675 &  -5.71094 & 1342216897 & sparse & 1342218697 & 1342218697 \\
20 & PN Mz 3 OFF & 244.28579 & -52.03330 & 1342251316 & sparse & 1342204046 & 1342204047 \\
21 & CTB37A-N ref & 258.24149 & -37.84571 & 1342251320 & sparse & 1342214511 & 1342214511 \\
22 & G349.7 ref & 259.06902 & -37.19761 & 1342251324 & sparse & 1342214511 & 1342214511 \\
23 & G357.7 ref & 264.50619 & -30.01860 & 1342251327 & sparse & 1342204367 & 1342204367 \\
24 & G357.7B-IRS & 264.61196 & -30.57159 & 1342251328 & mapping & 1342204367 & 1342204369 \\
\hline\hline
\end{tabular}
\label{tab_obsids}
\end{table*}

\subsection{Synthetic photometry from extended-source calibrated spectra}
\label{sec:synth}

To derive synthetic photometry from a spectrum we follow the approach explained in H17 and in \citet{Griffin_2013}. The total RSRF-weighted in-beam flux density from a source with spectral energy distribution $I_S(\nu)$ is
\begin{equation}
\bar S_S \left[ Jy \right] = \frac{\int_{\mathrm{passband}} I_S(\nu) \eta(\nu) R(\nu) \Omega(\nu) \mathrm{d}\nu}{\int_{\mathrm{passband}} \eta(\nu) R(\nu) \mathrm{d}\nu}.
\label{eq:is}
\end{equation}
Here $R(\nu)$ and $\eta(\nu)$ are the photometer spectral response function and the aperture efficiency for the passband. $\Omega(\nu)$ is the beam solid angle modelled with
\begin{equation}
\Omega(\nu) = \Omega(\nu_0) \left(\frac{\nu}{\nu_0}\right)^{2\gamma},
\end{equation}
where $\Omega(\nu_0)$ is the beam solid angle derived from Neptune and $\gamma = -0.85$, $\nu_0$ is the adopted passband central frequency. The Neptune derived beam solid angles at the band centres (250, 350, 500) \micron\ are $\Omega(\nu_0) = (469.35, 831.27, 1804.31)$ arcsec$^{2}$ (see H17).

A common convention in astronomy is to provide monochromatic flux densities or intensities at a particular central frequency $\nu_0$, assuming a source with a power law spectral shape: $I(\nu) \propto \nu^{-1}$. This convention is also used to calibrate the SPIRE photometer timelines. Hence, to convert $\bar S_S$ to monochromatic intensity $I_S(\nu_0)$ in [MJy/sr] for a source with $I(\nu) \propto \nu^{-1}$  we use
\begin{equation}
I_S(\nu_0) = \mathrm{KMonE}(\nu_0) \times \bar S_S,
\label{eq:synth}
\end{equation}
where the conversion factors $\mathrm{KMonE}(\nu_0)$ is
\begin{equation}
\mathrm{KMonE}(\nu_0) = \frac{\nu_0^{-1}\int_{\mathrm{passband}} \eta(\nu) R(\nu) \mathrm{d}\nu}{\int_{\mathrm{passband}} \nu^{-1} \eta(\nu) R(\nu) \Omega(\nu) \mathrm{d}\nu},
\end{equation}
and the corresponding values are (91.567, 51.665, 23.711) in units of [MJy/sr per Jy/beam] for the three photometer bands at (250, 350, 500) \micron. 

We use \autoref{eq:is} and \autoref{eq:synth} to derive the synthetic photometry of extended-source calibrated spectra $I_S(\nu)$ from the two co-aligned central detectors of the two FTS bands. The error on the synthetic photometry is calculated by substituting $I_S(\nu)$ in \autoref{eq:is} with $I_S(\nu) \pm \Delta I_S(\nu)$, where $\Delta I_S(\nu)$ is the standard error after averaging the different spectral scans in the pipeline (see \citealt{Fulton_2014} for details)\footnote{This framework is implemented in the \textit{Herschel} Interactive Processing Environment (HIPE) as a task \texttt{spireSynthPhotometry()}. The output of the task is the synthetic surface brightness values at 250, 350 and 500 \micron\ in MJy/sr, for a monochromatic fully extended source with $I(\nu) \propto  \nu^{-1}$.}.

The 250 and 500 \micron\ photometer bands are fully covered by the SSW and SLW spectra; however the 350 \micron\ band is mostly in SLW but a small fraction falls within SSW (see \autoref{fig:cal}). For a source with $I(\nu) \propto \nu^{-1}$, the underestimation of the synthetic photometry is $\sim 1\%$ and for a $\nu^{2}$ spectrum it is overestimated by $\sim 2\%$. These are within the overall calibration uncertainties and consequently we do not stitch together the SSW and SLW spectra before deriving the synthetic photometry at 350 \micron. 

\subsection{Comparison with the photometer}

For each of the 24 flat sources we derive synthetic photometry as described in \autoref{sec:synth}. The resulting values can be directly compared to the corresponding extended-source calibrated Photometer maps, by using a suitable aperture to take the average surface brightness. We use a square box aperture of 30\arcsec, which differs from the one used for the selection of extended and flat sources (sec.~\ref{sec:flats}). However, since we are averaging the surface brightness of flat extended sources then the choice of aperture is not important, as long as the size is comparable with the FTS beam. 

\begin{figure}
\centerline{\includegraphics[width=8cm]{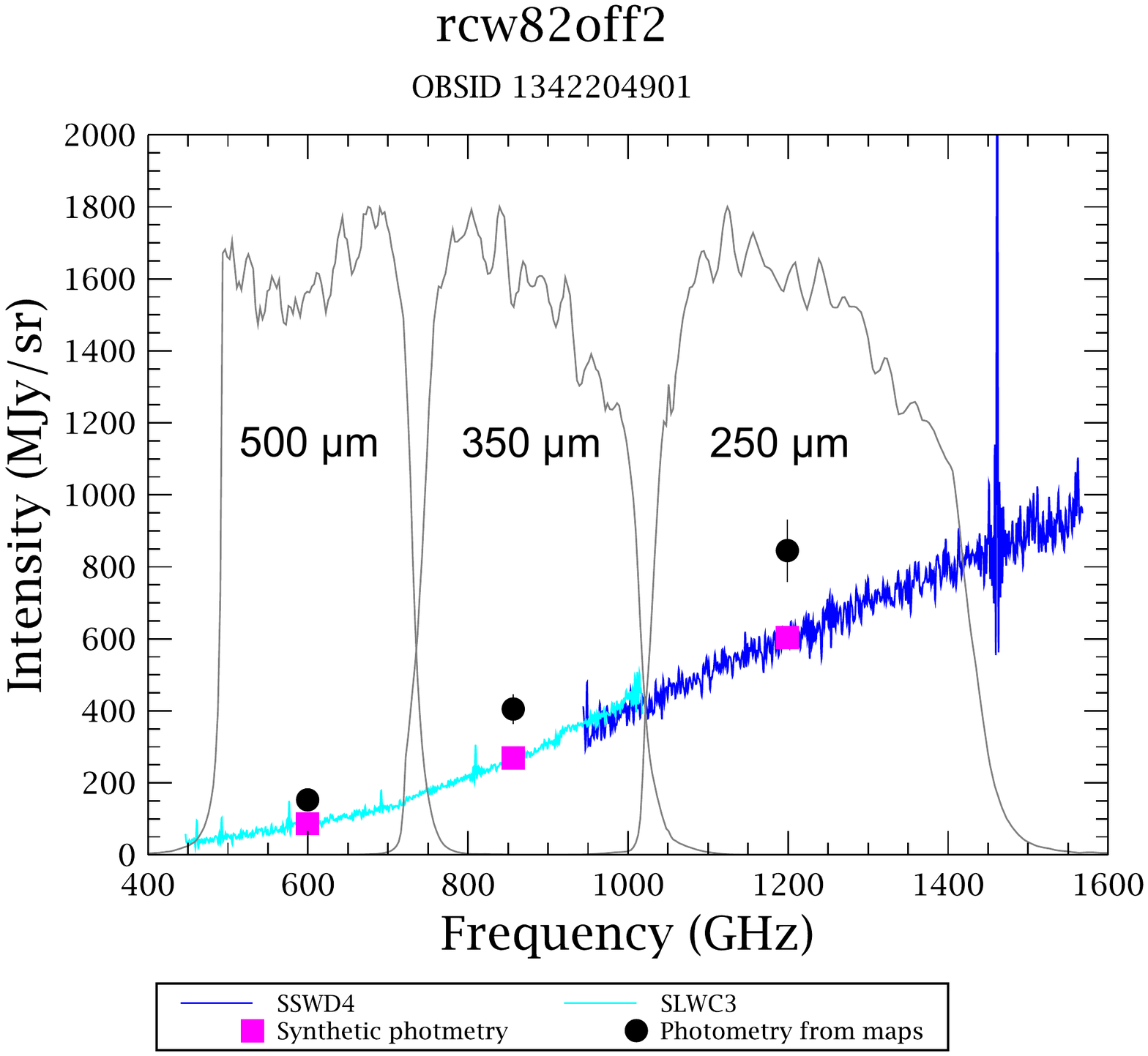}}
\caption{Comparison of the synthetic photometry from the extended-source calibrated spectra from version 13.1 of the pipeline and the surface brightness from photometer maps. The same source, RCW82off2, as in \autoref{fig:flat} is shown. The spectra are shown in blue for SSW and in cyan for SLW. The derived synthetic photometry points at the three photometer bands are shown as filled magenta squares. The error bars are smaller than the symbols and they include the errors from the scan-averaged spectra (see \citealt{Fulton_2014}). The photometer RSRFs are shown in grey, each one annotated with the band central wavelength. The average surface brightness values from photometer maps are shown as filled black circles. The error bars for the photometer points include the 10\% \textit{Planck}-to-SPIRE zero offset uncertainty and the standard deviation of the brightness distribution in the selected box. }
\label{fig:cal}
\end{figure}

Figure~\ref{fig:cal} shows the extended-source calibrated spectrum produced with version 13.1 of the FTS pipeline\footnote{Version 13.1 of the pipeline is the last one before the correction described in this paper was implemented.} for one of the flat sources (RCW82off2, ID8 in Table~\ref{tab_obsids}) and the derived synthetic photometry compared with the average surface brightness on photometer maps within the 30\arcsec\ box aperture. It is obvious that there is a significant offset between the synthetic photometry and the measured photometry in maps, with ratios of phot/spec $(1.38 \pm 0.10,\ 1.50 \pm 0.06,\ 1.77 \pm 0.20)$ at (250, 350, 500) \micron\ for this particular target. 

\begin{figure}
\includegraphics[width=8cm]{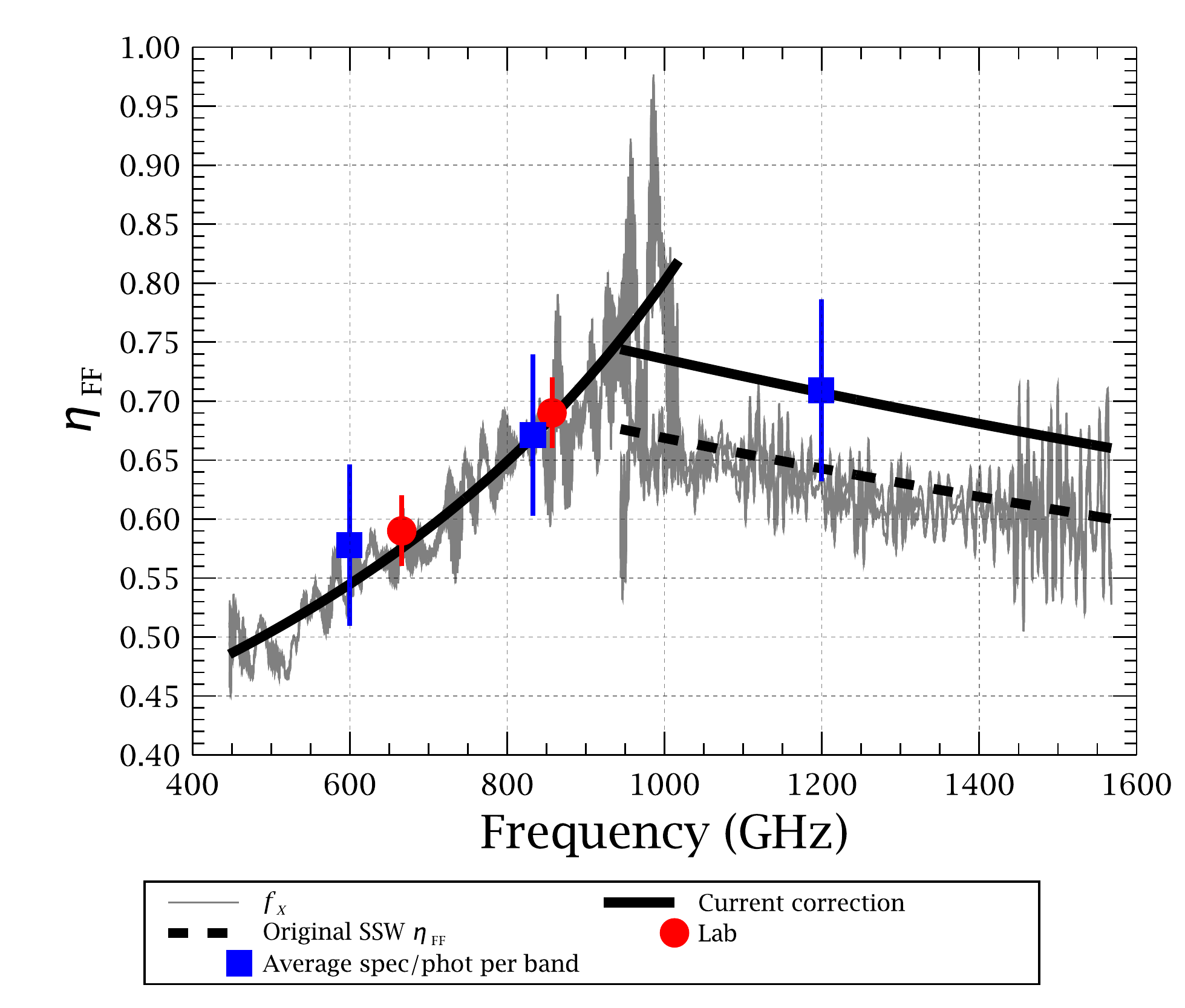}
\caption{Averaged ratios of the synthetic photometry versus the results from photometer maps for all 24 flat sources (filled blue squares), together with the far-field feedhorn efficiency (black lines, see \autoref{eq_7}) and the laboratory measurements from \citet{Chattopadhyay_2003} (filled red circles). The dashed line is the original \etaff\ for SSW as presented in \citet{Wu_2013}. The grey curves are the ratios $f_X = \etadiff\,I_S\, \Omega_\mathrm{beam}/S_S$ for all of the 24 flat sources (see \autoref{sec:etadiff}).}
\label{fig:etaff}
\end{figure}

The combined results for the averaged spec/phot ratio for each band for all 24 flat sources are shown as blue squares in \autoref{fig:etaff}. The errors bars for each point include the standard deviation of the aperture photometry, the 10\% uncertainty from the \textit{Planck} zero level offset and the error from the synthetic photometry. This figure unequivocally demonstrates that there is a systematic and significant discrepancy between the FTS and photometer extended-source calibrations. 

\subsection{The far-field feedhorn efficiency}
\label{sec:etaff}

The results shown in \autoref{fig:etaff} (as well as the example in \autoref{fig:cal}) indicate that in order to match the spectra with the photometry from extended-source calibrated maps we need to apply a correction. We consider the SPIRE Photometer extended-source calibration more straightforward than that of the the spectrometer: simple beam profile, uni-modal Gaussian beam and the beam solid angle is known down to $<1\%$uncertainty, and is consequently much more representative and robust. Moreover, the photometer maps are cross-calibrated with \textit{Planck}-HFI. Therefore the correction should be applied to the SPIRE FTS extended-source calibrated spectra.

The derived ratios, shown in \autoref{fig:etaff}, are a good match to the far-field feedhorn efficiency curve, \etaff. The correction, \etaff\ was introduced in empirical form in \citet{Wu_2013}, where it was linked with two other corrections: the diffraction loss predicted by the optics model, \etadiff\ \citep{Caldwell_2000} and the correction efficiency $\eta_c$, with  $\etaff = \eta_c/\etadiff$. As discussed in \citet{Wu_2013}, for point-like sources $\eta_c \approx 1$, while for extended sources $\eta_c \ll 1$ with the difference attributed to a combination of diffraction losses (\etadiff) and different response of the feedhorns and bolometers to a source filling the aperture and to that of a point source. 

The far-field feedhorn efficiency \etaff\ was measured by \citet{Chattopadhyay_2003} but only for the SLW band (the two laboratory measurements are shown as red circles in \autoref{fig:etaff}). The empirical \etaff\ from \citet{Wu_2013} is 10\% lower for SSW (shown as a dashed line in \autoref{fig:etaff}) with respect to the measured ratio at 250\,\micron. This 10\% is within the uncertainty of the 250\,\micron\ average ratio, however,  the original empirical \etaff\  would introduce a significant discontinuity in the overlap region of the two FTS bands (944--1018\,GHz). In order to avoid this inconsistency, \etaff\ was rescaled by 10\% for SSW, so that it matches the 250\,\micron\ ratio and also avoids the discontinuity. It is irrelevant to attribute this 10\% offset to any parameter in the optical model (\etadiff, \citealt{Caldwell_2000}). The most likely interpretation is that some unknown effects in the complicated feedhorn-coupled system lead to a different response for fully extended sources only for SSW, which leads to $\eta_c = 1.1$ for SSW, while for SLW $\eta_c = 1$.

\begin{figure}
\includegraphics[width=8cm]{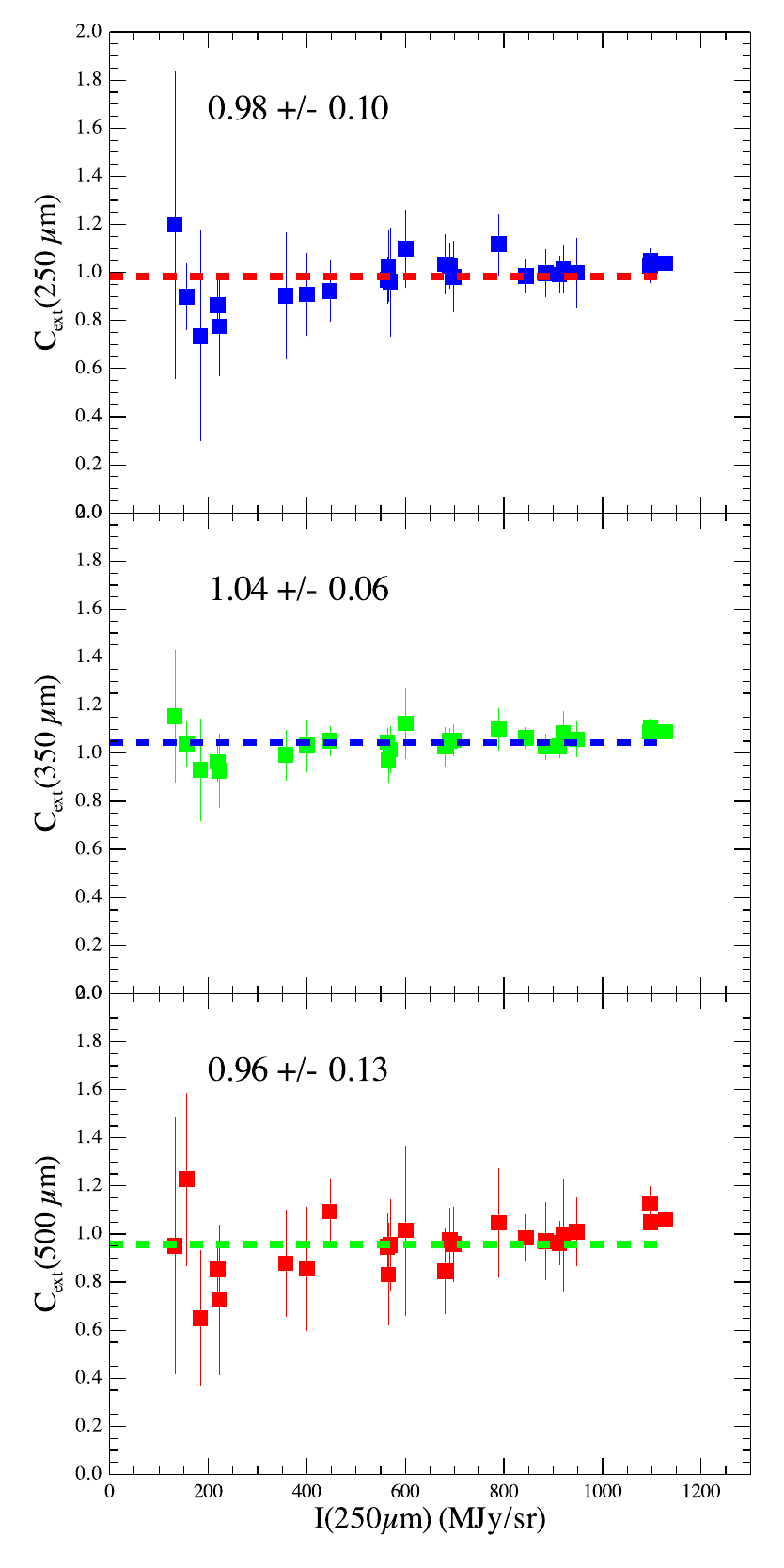}
\caption{$C_\mathrm{ext}(\nu_0) = I_\mathrm{phot}(\nu_0)/I_\mathrm{spec}(\nu_0)$ as a function of $\bar I_{250\micron}$ at 250 \micron (top), at 350 \micron\ (middle) and 500 \micron\ (bottom) for the 24 flat sources. The mean (shown as a dashed line) and the standard deviation for each band are annotated in each panel: from top to bottom, 250, 350 and 500 \micron.}
\label{fig:ratio}
\end{figure}

In practice, due to implementation considerations, we use the following empirical approximation based on the \etaff\ curves shown in Figure~4 in \citet{Wu_2013}, with SSW rescaled by 10\%: 
\begin{align}
\label{eq_7}
\mathrm{SLW}: 1/\etaff &=  2.7172 - 1.47\times 10^{-3}\nu,\\ \nonumber
\mathrm{SSW}: 1/\etaff &=  1.0857 + 2.737\times 10^{-4}\nu,
\end{align}
where $\nu$ is the frequency in GHz. The two curves are shown in \autoref{fig:etaff}. And the corrected intensities are
\begin{equation}
I^{\prime}_\mathrm{ext} = I_\mathrm{ext}/\etaff, 
\label{eq_etaff}
\end{equation}
where $I_\mathrm{ext}$ is the extended-source calibrated spectrum from \citet{Swinyard_2014} calibration (see also \autoref{eq:iext}). Performing the same comparison for $I^{\prime}_\mathrm{ext}$ with the extended-calibrated maps from the photometer for the 24 flat sources, we obtain the ratios as shown in \autoref{fig:ratio}. On average we see a good agreement at a level of 2--4 \%, comparable to that found for the point-source calibration in \citet{Hopwood_2015}.

\section{Converting to point-source calibration}
\label{sec:etadiff}

For an extended source on the sky $I_S(\theta,\phi)$, the measured flux density is 
\begin{equation}
S_S(\nu) = \eta \oint_{4\pi} P(\theta,\phi)\ I_S(\theta,\phi) \mathrm{d}\Omega,
\end{equation}
where $P(\theta,\phi)$ is the normalised beam profile and $\eta$ represents all angle-independent efficiency factors that affect the system gain. The integration is over a region subtended by the source. 

For a spatially flat source, $I(\theta,\phi) = I_S(\nu)$ = constant, and assuming that the source is much more extended than the beam, we can write
\begin{equation}
S_S(\nu) = \eta \times I_S(\nu) \times \Omega_\mathrm{beam}(\nu),
\label{eq:check}
\end{equation}
where $\Omega_\mathrm{beam}(\nu) = \oint_{4\pi} P(\theta,\phi) \mathrm{d}\Omega$ is the main beam solid angle.

Equation~\ref{eq:check} should be valid for any instrument. And it is indeed the case for the SPIRE Photometer, where the conversion from point-source to extended-source calibrated maps can be achieved by multiplication with $K_\mathrm{PtoE}(\nu) \equiv \Omega_\mathrm{pip}$, where $\Omega_\mathrm{pip}$ is the beam solid angle used in the data processing pipeline (see H17 for more details). The gain and aperture corrections already incorporated in the point-source calibrated timelines in the data processing pipeline.% beam solid angle $\Omega_\mathrm{pip}$, respectively.

\begin{figure*}
\centerline{
\includegraphics[width=0.45\textwidth]{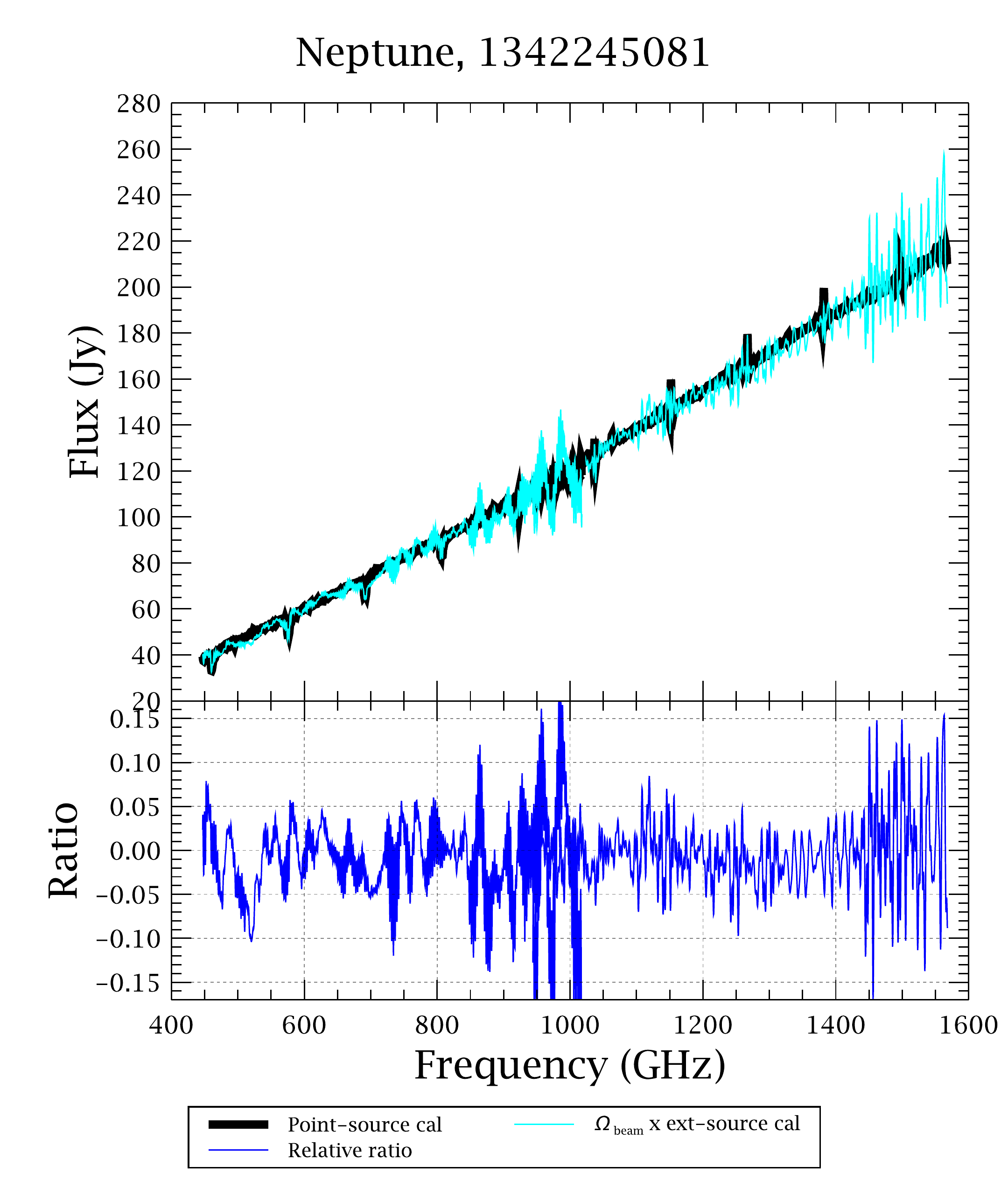}
\includegraphics[width=0.45\textwidth]{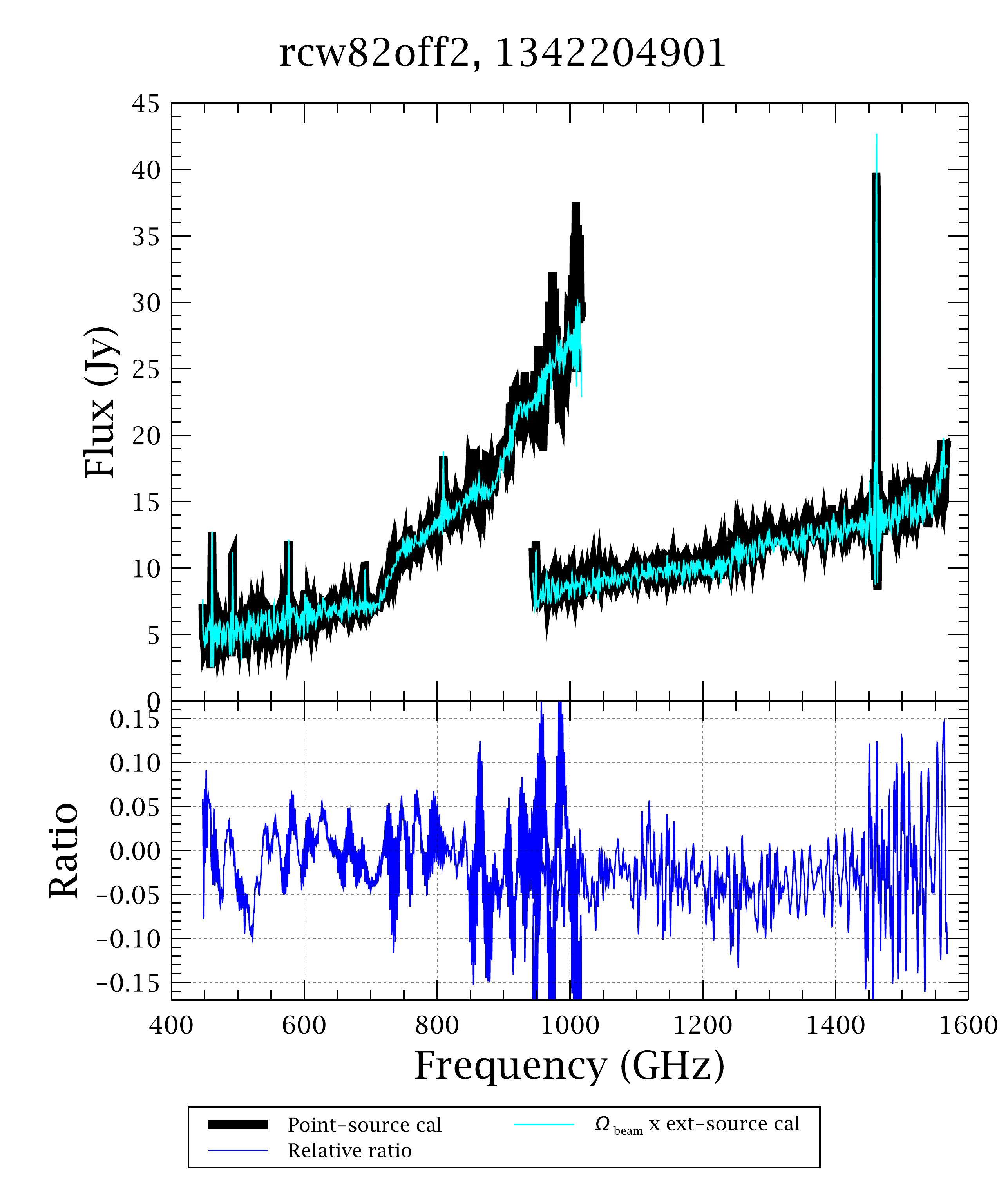}}
\caption{Left panel: comparison of Neptune pipeline derived point-source calibrated flux density $S_S(\nu)$ in Jy (thick black line) with the flux density derived from the extended-source calibrated intensity $S^{\prime}_S = \etadiff \times I^{\prime}_S(\nu)\times \Omega_\mathrm{beam}$ (cyan), i.e. \autoref{eq:check}. The relative ratio of $S^{\prime}_S/S_S$ is shown in the bottom panel. The overall agreement, in the less noisy parts of the two bands, is within 5\%. Right panel: the same comparison for a fully extended source.}
\label{fig:check}
\end{figure*}

The validity of \autoref{eq:check} for the corrected extended-source calibrated spectra is demonstrated in \autoref{fig:check} for a point source (Neptune) and an extended source from the sample of 24 spatially flat sources. In this case, the efficiency factor $\eta$ is actually the diffraction loss correction, \etadiff\ as derived by \citet{Caldwell_2000}, using a simple optics model, incorporating the telescope secondary mirror and mirrors support structures. For a point source on axis \etadiff\ is of the order of 75\%. We see that  Eq.~\ref{eq:check} is fulfilled at a level of $\pm 5\%$, if we exclude noisier regions close to the band edges (Figure~\ref{fig:check}, bottom panels).

The noise that appears in the point-source converted spectra in \autoref{fig:check} (cyan curves) reflects the small-scale characteristics of $R_\mathrm{tel}$ that are inherently present in $I^{\prime}_\mathrm{ext}$. The original point-source calibrated spectrum of Neptune (\autoref{fig:check}, left) has much less noise because the point-source calibration is based on the smooth featureless model spectrum of Uranus and consequently $C_\mathrm{point}$ accounts for those small-scale features of $R_\mathrm{tel}$. Therefore, the pipeline-provided point-source calibrated spectra are better products and they should be used, rather than converting the extended-source calibration with \autoref{eq:check}.

Interestingly, the missing correction for the old calibration of the FTS extended-source spectra is obvious, if we construct the ratio of the left-hand and right-hand side of \autoref{eq:check}, i.e. $f_X = \etadiff\,I_S\, \Omega_\mathrm{beam}/S_S$. This ratio should be one if \autoref{eq:check} is valid, but as shown in \autoref{fig:etaff}, the grey curves, which are the derived $f_X$ for all 24 flat sources with the old calibration, match well with the empirical \etaff\ instead.

\section{Practical considerations}
\label{sec:practical}

All extended-source calibrated spectra, regardless of the observing mode and the spectral resolution, are corrected for the missing far-field feedhorn efficiency (Equation~\ref{eq_etaff}). Using those for analysis of extended sources is straightforward: measuring lines and the continuum, with results in the corresponding units of W m$^{-2}$ sr$^{-1}$. A large fraction of the sources observed with the FTS, however, are neither point-like nor fully extended, we call them semi-extended sources. The framework for correcting the spectra for this class of targets is presented in \citet{Wu_2013} and implemented in HIPE as an interactive tool -- the \textsc{semiExtendedCorrector} (SECT). There are two possible ways to derive a correction for the source size (and/or a possible pointing offset): starting from an extended-source or from a point-source calibrated spectrum (see \citealt{Wu_2013}, Eq. 14). The SECT implementation in HIPE follows the procedure starting from a point-source calibrated spectrum. As the point-source calibration is not affected by the far-field feedhorn efficiency correction, described in Section~\ref{sec:etaff}, so there should not be any changes in the SECT-corrected spectra.

In cases when there is a point source embedded in extended emission, then the background subtraction should be performed using the point-source calibrated spectra, regardless of the fact that the background may be fully extended in the beam. If you perform the background subtraction using $I^{\prime}_\mathrm{ext}$, then you cannot any longer use $C_\mathrm{point}$ to convert the background subtracted spectrum to a point-source calibrated one. Instead, you have to use Equation~\ref{eq:check}, and as explained in Section~\ref{sec:etadiff}, this will introduce unnecessary noise in the final spectrum.

The same consideration is applicable for semi-extended sources, where the first step before the correction should be the background subtraction and then proceeding with SECT, both steps should be performed on point-source calibrated spectra.

Careful assessment of the source extension is always necessary, because in some cases the source may fall in the extended source category in continuum emission but semi-extended or point-like in a particular line transition. This will dictate which calibration to use and what corrections to apply to the line flux measurements.

Finally, if for some reason one needs to recover the spectrum with the original calibration  following \citet{Swinyard_2014}, then  $C_\mathrm{point}$\footnote{$C_\mathrm{point}$ is available as a calibration table within the SPIRE calibration context (see H17 and appendix \ref{data_products}).} and the point-source calibrated spectrum can be used: $I_\mathrm{ext} = S_S/C_\mathrm{point}$.

\section{Implications for SPIRE FTS users and already published results}
\label{sec:impact}

The significant correction for the extended-source calibration scheme presented by this work, was implemented as of HIPE version 14.1, and has already been described in H17 since Feb 2017. All analysis based on extended-source calibrated FTS spectra, produced prior to that version, will be affected by the significant and systematic shortfall of the old calibration. Any integrated line intensity or continuum measurements will be underestimated by a factor of 1.3--2 and using them to derive physical conditions in objects will be subject to corresponding systematic errors.

\begin{figure}
\centerline{
\includegraphics[width=8cm]{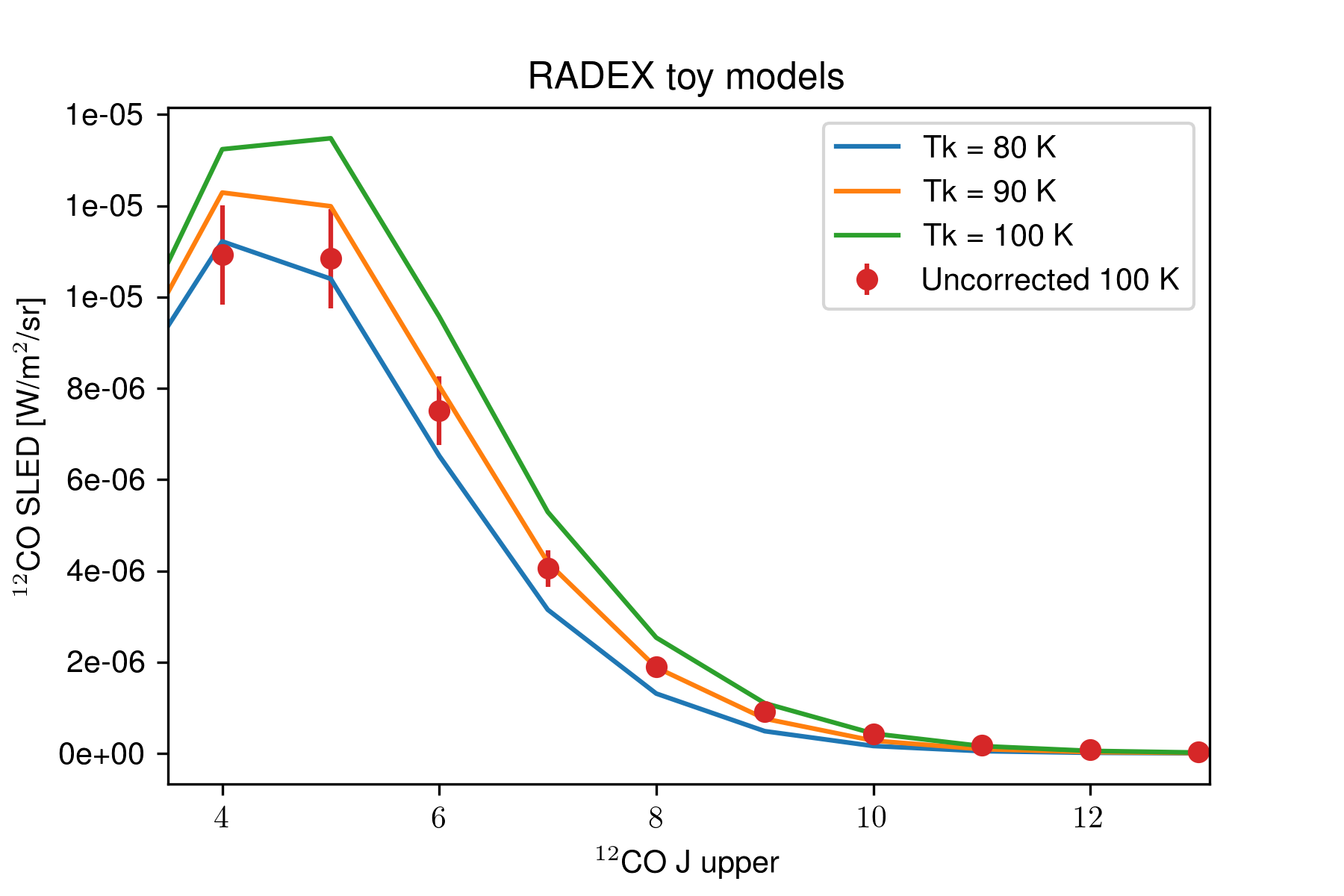}
}
\caption{$^{12}$CO spectral line energy distribution model from RADEX \citep{radex} for an emitting region, assuming $n(\mathrm{H}_2) = 6.3\times10^3$ cm$^{-3}$, column density of $10^{16}$ cm$^{-2}$ and kinetic temperatures of 100 K (green curve), 90 K (orange) and 80 K (blue). The 100 K SLED is multiplied by \etaff\ and the new uncorrected SLED is shown with red points with error bars assuming a conservative 10\% uncertainty in line flux measurements.}
\label{fig:sled}
\end{figure}

To illustrate the magnitude of the deviations on the derived physical characteristics with the old calibration, we performed a simple simulation using RADEX \citep{radex}. We modelled the spectral line energy distribution (SLED) of the $^{12}$CO lines from an emitting region with molecular hydrogen density $n(\mathrm{H}_2) = 6.3\times10^3$ cm$^{-3}$, column density of $10^{16}$ cm$^{-2}$ and kinetic temperatures $T_\mathrm{kin}$ of 100, 90 and 80 K. The predicted line fluxes for the three temperatures in the SPIRE FTS bands are shown in Figure~\ref{fig:sled} as green, orange and blue curves respectively. 

If we observe a region with $T_\mathrm{kin} = 100$ K, but we use the old calibration, then the measured $^{12}$CO lines (the green curve ) will be underestimated by a factor of \etaff; these are shown in \autoref{fig:sled} as red points with 10\% measurement errors. Obviously the red points do not match the RADEX models with $T_\mathrm{kin}=100$ K, they are at least 2-3 $\sigma$ away from the correct input model for lines with upper-J $\leq 8$. While models with $T_\mathrm{kin}$ between 85 and 90 K are much closer to the ``measurements'' and consequently the derived temperature from the red points will be significantly underestimated. 

Using the old calibration for studies based on line-to-line or line-to-continuum measurement will not be significantly biased for SSW, because the variation of \etaff\ with frequency within the band is small. However, the variation across SLW is significant and in this case using uncorrected data will lead to the incorrect results.

The \etaff\ correction to extended-source calibrated spectra results in new values for the frequency dependent additive continuum offsets and FTS sensitivity estimates (see \citealt{Hopwood_2015}). The new offsets and sensitivities are presented in H17 and their tabulation is available in the \herschel\ legacy repository as Ancillary Data Products\footnote{See Appendix~\ref{data_products} with a list of URLs for the data products.}. 

The correction with \etaff\ also introduces a new source of uncertainty to the overall calibration error budget for extended sources. The two measurement points for \etaff\ in SLW band have errors of 3\% \citep{Chattopadhyay_2003}, and we assume the same error is applicable for the SSW band. Therefore the overall calibration accuracy budget for extended-source calibration will have to incorporate the 3\% statistical uncertainty on \etaff. As the correction is semi-empirical and based on cross-calibration with the SPIRE Photometer, the more conservative estimate of the overall uncertainty is of the order of 10\%, to match the uncertainties on the derived photometry ratios (\autoref{fig:ratio}).

\section{Conclusions}
\label{sec:conclusions}

We introduce a correction to the SPIRE FTS calibration for the far-field feedhorn efficiency, \etaff. This brings the cross-calibration between extended-source calibrated data for the spectrometer and photometer in agreement at a 2--4\% level for fully extended and spatially flat sources. With this correction, the FTS point-source and extended-source calibration schemes are now self-consistent and can be linked together using the beam solid angle and a gain correction for the diffraction losses.

All SPIRE FTS extended-source calibrated products (spectra or spectral maps) in the \herschel\ Science Archive, processed with pipeline version 14.1 have already been corrected for \etaff. Spectra processed with earlier versions are significantly underestimated and consequently the results derived with the old calibration should be critically revised. It is important to note, that while the correction is close to a constant factor for the SSW band, this is not the case for SLW. Hence, even relative line-to-line or line-to-continuum analysis for SLW is affected. 

We have not discussed any possible reason as to why the far-field feedhorn efficiency was not naturally incorporated in the extended-source calibration scheme. With \textit{Herschel} no longer operational, it is not possible to take new measurements in order to check any hypothesis. We can only speculate about possible causes. One plausible reason is that the FTS beam, which was only measured out to a radial distance of 45\arcsec, compared to the 700\arcsec\ for the Photometer, has an important fraction of the power distributed at larger distances, or in the side-lobes. Another possibility could be that the coupling of the two instruments to extended sources, viewed through the telescope, differs in an unknown manner such as small residual misalignment. Both these hypotheses could play a part in \etaff\ not being naturally incorporated into then extended-source calibration. The bottom line, however, is that with this correction the FTS calibration is now self-consistent and the cross-calibration with the SPIRE Photometer is in good agreement.

Ground based measurements of lines or continuum, in frequency ranges that overlap with the large spectral coverage of the FTS, may provide further insights on the correctness of the extended-source calibration, although the direct comparison will not be straightforward due to the complications in observing very extended emission with ground-based telescopes.

\section*{Acknowledgements}

The authors wish to thank the referee for their useful comments that helped improve the paper, as well as J. Kamenetzky, C. Wilson, D. Teyssier, K. Rygl, E. Puga and K. Exter for valuable discussions.\\
SPIRE has been developed by a consortium of institutes led by Cardiff Univ. (UK) and including: Univ. Lethbridge (Canada); NAOC (China); CEA, LAM (France); IFSI, Univ. Padua (Italy); IAC (Spain); Stockholm Observatory (Sweden); Imperial College London, RAL, UCL-MSSL, UKATC, Univ. Sussex (UK); and Caltech, JPL, NHSC, Univ. Colorado (USA). This development has been supported by national funding agencies: CSA (Canada); NAOC (China); CEA, CNES, CNRS (France); ASI (Italy); MCINN (Spain); SNSB (Sweden); STFC, UKSA (UK); and NASA (USA). This research is supported in part by the Canadian Space Agency (CSA) and the Natural Sciences and Engineering Research Council of Canada (NSERC). \\
Most of the data processing and analysis in this paper was performed in the \textit{Herschel} Interactive Processing Environment (HIPE, \citealt{hipe}).

\herschel\ is an ESA space observatory with science instruments provided by European-led Principal Investigator consortia and with important participation from NASA.

\bibliography{ftsbiblio}{}

\appendix
\section{Available data products}
\label{data_products}

Many useful calibration tables are available in the \herschel\ Legacy Area at \url{http://archives.esac.esa.int/hsa/legacy}. Here we only list those with relevance to the current paper.

\begin{itemize}
\item \textbf{Planetary models}: \\ Models for the primary calibrators (Uranus and Neptune) are available at \url{http://archives.esac.esa.int/hsa/legacy/ADP/PlanetaryModels/}
\item \textbf{FTS Sensitivity curves and additive continuum offsets}: \\ The curves derived from the updated calibration are available at \url{http://archives.esac.esa.int/hsa/legacy/ADP/SPIRE/SPIRE-S_sensitivity_offset/}
\item \textbf{Diffraction loss curves}: \\ The correction \etadiff\ as presented in \citet{Wu_2013}, and based on the optics model from \citet{Caldwell_2000} is available at \url{http://archives.esac.esa.int/hsa/legacy/ADP/SPIRE/SPIRE_Diffraction_loss/}
\item \textbf{SPIRE Photometer RSRFs}: \\ The Relative Spectral Response Functions $R(\nu)$ and the aperture efficiencies $\eta(\nu) $ are available at \url{http://archives.esac.esa.int/hsa/legacy/ADP/SPIRE/SPIRE-P_filter_curves/}
\item \textbf{SPIRE Calibration Tree}: \\The last one (\textsc{spire\_cal\_14\_3}) as well as previous version of the calibration tables are available as Java archive files (jar) at 
\url{http://archives.esac.esa.int/hsa/legacy/cal/SPIRE/user/}
\end{itemize}

\label{lastpage}

\end{document}